\documentclass[twocolumn]{aastex62}

\usepackage{latexsym}
\usepackage{xcolor}
\usepackage{amsmath}
\usepackage{amssymb}
\usepackage{graphicx}
\usepackage{aas_macros}

\definecolor{texas}{HTML}{BF5700}

\graphicspath{{./}{figures/}}

\submitjournal{ApJ}

\shorttitle{Dwarf galaxy formation with and without streaming velocities}
\shortauthors{Schauer et al.}

\begin{document}

\title{Dwarf galaxy formation with and without dark matter-baryon streaming velocities}

\correspondingauthor{Anna T. P. Schauer}
\email{anna.schauer@utexas.edu}

\author[0000-0002-2220-8086]{Anna T. P. Schauer}
\affiliation{Department of Astronomy, 
University of Texas at Austin,
TX 78712, USA}

\author[0000-0002-9604-343X]{Michael Boylan-Kolchin}
\affiliation{Department of Astronomy,
University of Texas at Austin,
TX 78712, USA} 

\author{Katelyn Colston}
\affiliation{Department of Astronomy,
University of Texas at Austin,
TX 78712, USA} 

\author[0000-0003-4394-6085]{Omid Sameie}
\affiliation{Department of Astronomy,
University of Texas at Austin,
TX 78712, USA} 

\author[0000-0003-0212-2979]{Volker Bromm}
\affiliation{Department of Astronomy,
University of Texas at Austin,
TX 78712, USA} 

\author[0000-0003-4298-5082]{James S. Bullock}
\affiliation{Department of Physics and Astronomy,
University of California, Irvine,
CA 92697, USA }

\author[0000-0003-0603-8942]{Andrew Wetzel}
\affiliation{Department of Physics and Astronomy, 
University of California, Davis, 
CA 95616, USA}

\begin{abstract}
We study how supersonic streaming velocities of baryons relative to dark matter --- a large-scale effect imprinted at recombination and coherent over $\sim 3$ Mpc scales --- affects the formation of dwarf galaxies at $z \gtrsim 5$. We perform cosmological hydrodynamic simulations, including and excluding streaming velocities,  
in regions centered on halos with
$M_{\rm vir}(z=0) \approx 10^{10}$\,M$_{\odot}$; the simulations are part of the Feedback In Realistic Environments (FIRE) project and run with FIRE-3 physics. 
Our simulations comprise many thousands of systems with halo masses between $M_{\rm vir} = 2\times10^{5}$\,M$_{\odot}$ and $2\times10^9$\,M$_{\odot}$ in the redshift range $z=20-5$. A few hundred of these galaxies form stars and have stellar masses ranging from 100 to $10^7\,$M$_{\odot}$. While star formation is globally delayed 
by approximately 50\,Myr in the streaming relative to non-streaming simulations 
and the number of luminous galaxies is correspondingly suppressed at high redshift in the streaming runs, these effects decay with time. By $z=5$, the properties of the simulated galaxies are nearly identical in the streaming versus non-streaming runs, indicating that any effects of streaming velocities on the properties of galaxies at the mass scale of classical dwarfs and larger do not persist to $z=0$. 
\end{abstract}

\keywords{early universe --- dark ages, reionization, first stars ---
galaxies: dwarf --- galaxies: high-redshift}

\section{Introduction}
The high-redshift Universe is a transformative time in cosmic history studied by numerous numerical simulations, leveraging a broad array of computational methodologies (e.g. \citealt{bromm02b,yoshida03,clark11,wise12,hirano18}). 
With structure formation just starting out, dark matter halos are generally much smaller and very filamentary \citep{sasaki14}. 
Without metals, gas cooling proceeds via hydrogen, helium and their isotopes and molecules, with molecular hydrogen being the dominant coolant before metal enrichment takes place \citep{glov13}.

However, one significant effect that is important for the infall of baryons into low-mass halos at high-redshift 
was discovered less than 15 years ago \citep{th10} and has not been explored in nearly the same detail: there are ``streaming velocities'' of baryons relative to dark matter that are sourced by acoustic oscillations and imprinted at recombination ($z\sim 1100$). This offset velocity is Gaussian, and the speed of baryons relative to dark matter at any point therefore follows a Maxwell-Boltzmann distribution across large scales in the Universe that is coherent on scales smaller than $\sim 3$~Mpc (comoving) and correlated over scales comparable to the acoustic horizon of $\sim 150$~comoving Mpc \citep[see, e.g.,][for a review]{fialkov14}. 

The rms amplitude of this streaming velocity $v_{\rm bc}$ between baryons and cold dark matter is $\sigma_{\rm bc} \approx 30$\,km\,s$^{-1}$ at recombination. Like all peculiar velocities, it decays as the Universe expands as $v_{\rm bc}(z) \propto v_{\rm bc}(z=1090)\,(1+z)$, but it still plays a significant role for the formation of the first stars. 
Studies have shown that minihalos, the formation sites of Population~III (Pop~III) stars, 
have a smaller baryon content when located in a region of the Universe with a significant streaming velocity \citep{Naoz11,naoz12,schauer19,conaboy22}. 
Even more important, the minimum halo mass for first star formation increases \citep{greif11,stacy11,hirano18,schauer19}, an effect that is even stronger than the presence of H$_2$ dissociating Lyman-Werner radiation \citep{machacek01, johnson08, hirano15}. 
Furthermore, streaming velocities might be associated with direct collapse black hole formation \citep{hirano17, schauer17}. 

On a global scale, streaming velocities significantly influence the 21\,cm signal of neutral hydrogen during the Dark Ages before reionization \citep{visbal12,fialkov18,schauer19b,munoz22}. 
The redshift of reionization itself can vary by $\Delta z=0.05-0.5$, depending on the model for X-ray heating and detailed timeline of ionizing sources \citep{park20}. 
The signature of the streaming velocity is, however, lost at low redshift: both the power spectrum \citep{yoo13} and the three-point correlation function \citep{slepian18} of luminous red galaxies at redshift $0.4-0.7$ allow for only a small imprint of the streaming velocity in large-scale structure.

While the effect of streaming velocities on large scales is 
likely unmeasurable at low redshifts, it is possible that an archaeological imprint might remain in galaxies: if streaming velocities significantly affect early star formation in low-mass halos, present-day dwarf galaxies might retain this signature in their stellar populations. With increasingly high resolution simulations, it has been found that dwarf galaxies assemble early \citep{ricotti05,jeon17,fitts18}. 
The ability, therefore, of even smaller halos to form stars is critical for the stellar mass content in dwarf galaxies at the epoch of reionization. Investigating how a high-redshift effect influences the first galaxies is a natural first step in connecting physics of the high-redshift Universe with small galaxies in reach of observations. While dwarf galaxies have been studied extensively 
for many decades with both observations and simulations, there remain important open questions, such as the too-big-to fail problem \citep{mbk11}, 
the diversity of rotation curves \citep[e.g.][]{oman15,santossantos20}, 
or the planes that satellites are found in (\citealt{mbk21,pawlowski21}, for a review see \citealt{sales22}).  

In this study, we investigate for the first time the role of streaming velocities in the formation of dwarf galaxies at the epoch of reionization.  
Specifically, we directly compare global properties such as star formation history, halo mass function, or metal enrichment in three FIRE-3 dwarf galaxy simulations with and without steaming velocities. We further investigate the halo properties in all six simulations between redshifts $z=20$ and $z=5$. Our paper is structured as follows: We give an overview of the methodology (the simulations, the implementation of the streaming velocity, our halo selection criteria) in Section \ref{sec:methods}. 
We present the  results from large scales to small scales and from a global perspective to individual halos in Section \ref{sec:results}, before concluding and discussing caveats in Section \ref{sec:conclusions}. 
\section{Methodology} 
\label{sec:methods}
\subsection{Simulations}
We select three dwarf galaxies from the FIRE \citep{hopkins14,hopkins18} simulation suite with the latest FIRE-3 physics implemented \citep{hopkins22}: 
m10a, m10i, and m10m (see for example \citealt{fitts18}, \citealt{sameie22} for previous studies on these galaxies).  
These are high-resolution zoom-in simulations centered on individual halos of $M_{\rm halo}(z=0)=10^{10}$\,M$_\odot$ --- hosts of $M_{\star}(z=0) \sim 10^{6}$~M$_{\odot}$ galaxies --- in parent volumes of (25 comoving Mpc$/h)^3$. We focus on the high-redshift evolution of the galaxies between first star formation ($z=22-13$) and $z=5$. 

The FIRE-3 simulations are run with an updated version of the code GIZMO 
\citep{hopkins15}, and use the mesh-free finite mass (MFM) Lagrangian Godunov method. Mass, energy and momentum are conserved while the spatial scales adapt to resolve the simulation in high resolution. 
We work with a $\Lambda$CDM cosmology and the WMAP parameters from its 7-year data release \citep{komatsu11} with $h=0.71$, 
$\Omega_\mathrm{m} = 0.296 = 1- \Omega_\Lambda$, and $\Omega_\mathrm{b} = 0.0449$. 
These parameters differ slightly from the most recent results from the Planck collaboration 
\citep{planck20}; these differences are unimportant for our purposes.

While the numerical implementation remains largely unchanged from FIRE-2 to FIRE-3, multiple improvements have been made to the code. These updates include a state-of-the-art ionizing UV background that follows \cite{fg20}, explicit treatment of cooling in low-temperature, dense ($T<10^4$\,K and $n > 1$\,cm$^{-3}$) gas, and density-independent star formation criteria. The metallicity floor present in previous versions of FIRE has also effectively been removed in FIRE-3, though primordial chemistry is still not treated explicitly, meaning stars formed out of primordial gas follow a normal IMF with standard yields.
For all changes, we refer the reader to \cite{hopkins22}.

Our simulations are run with in high-resolution, with a dark matter particle mass of $M_\mathrm{DM} \sim 2460$\,M$_\odot$ and an average gas cell mass of $M_\mathrm{gas} \sim 500$\,M$_\odot$. 
The initial conditions are created at redshift $z=125$, and snapshots are written out at $z=50, 40, 30$, followed by outputs with $\Delta z=1$ until $z=20$, and then separated by a timescale of $\approx 10$\,Myr, resulting in 115 snapshots in total down to $z=5$.


\subsection{Inclusion of streaming velocities}
To rerun the same dwarf galaxy simulations m10a, m10i and m10m, with the inclusion of streaming velocities, we use the 
commonly employed ``baryons-trace-dark matter'' (BTD) approximation \citep{hirano18,schauer21}. 
Streaming velocities are coherent on the Silk damping scale \citep{silk68}, with a coherence length of $\sim$3\,cMpc \citep{th10}. 
The high-resolution cut-out regions for the galaxy simulation have a size of maximally 1.5\,cMpc/$h$ and are therefore smaller than the coherence length. 
Under the BTD approximation, we can assume that the velocity offset is constant within the cut-out region. 
At the redshift of 
initialization, $z=125$, we therefore add an additional constant velocity to all gas particles. 
We specifically add the velocity in the $x$-direction, which is an arbitrary choice and does not 
influence the results. 
Even though the BTD approximation artificially enhances the gas power spectrum by 
neglecting smoothing of the gas distribution between recombination and the redshift of 
initialization, the effects on a simulation are minor \citep{park20}. 

The streaming velocity value is chosen to be 1.945~$\sigma_\mathrm{bc}$, 
corresponding to 6.75\,km\,s$^{-1}$ at redshift $z=125$. 
A streaming velocity of 1.945~$\sigma_\mathrm{bc}$ or higher is present in 
1\% of the volume of the Universe. We therefore focus on a rare, but not extremely rare, region of the Universe with a large streaming velocity to maximize the effects.\footnote{For example, the CEERS JWST survey targets an area of 100~arcmin$^2$ \citep{ceers_prop} in the high-redshift Universe, corresponding to $\sim 800$\,cMpc$^3$ at $z=10$. The probability of finding at least one region with a streaming velocity of $v_\mathrm{bc} \ge 1.945$~$\sigma_\mathrm{bc}$ in this volume is more than 50\% at redshift $z=10$.} Simulations that include streaming velocities are labeled ``m10a-vbc'', etc., where vbc stands for velocity between baryons and cold dark matter, as commonly used in the literature.

\subsection{Halo Selection}
Halos are identified using the dark matter halo finder ROCKSTAR \citep{behroozi13}. 
We consider all halos at redshift $z=5$ that have a mass of at least $10^5$\,M$_\odot$, corresponding to at least 40 dark matter particles.
We also limit our halos to well-resolved ones with at least 99\% of the dark matter mass comprised of high-resolution dark matter particles (some halos at the edges of the high-resolution zoom-in region have significant fractions of low-resolution dark matter, and are excluded from this analysis). 

We categorize halos in main halos and subhalos: if the center of a halo lies within the virial radius of a more massive halo, it is a subhalo, otherwise it is a main halo. We compute the virial mass and radius of a halo relative to an overdensity of 200 times the cosmic background density at the given redshift. Gas and stars are attributed to a halo, if they lie within its virial radius. Stars further have to be kinematically bound to the halo by having a relative velocity with respect to the halo center of less than two times the escape velocity. Each star particle is at most associated with one halo, and we attribute it to the smaller halo (subhalo) if the aforementioned conditions are fulfilled by more than one halo. We count a halo as luminous / star forming if it hosts at least one star particle (so $M_\star\ge430$\,M$_\odot$).

\section{Results}
\label{sec:results}
\begin{figure*}
\centering
\includegraphics[width=\textwidth]{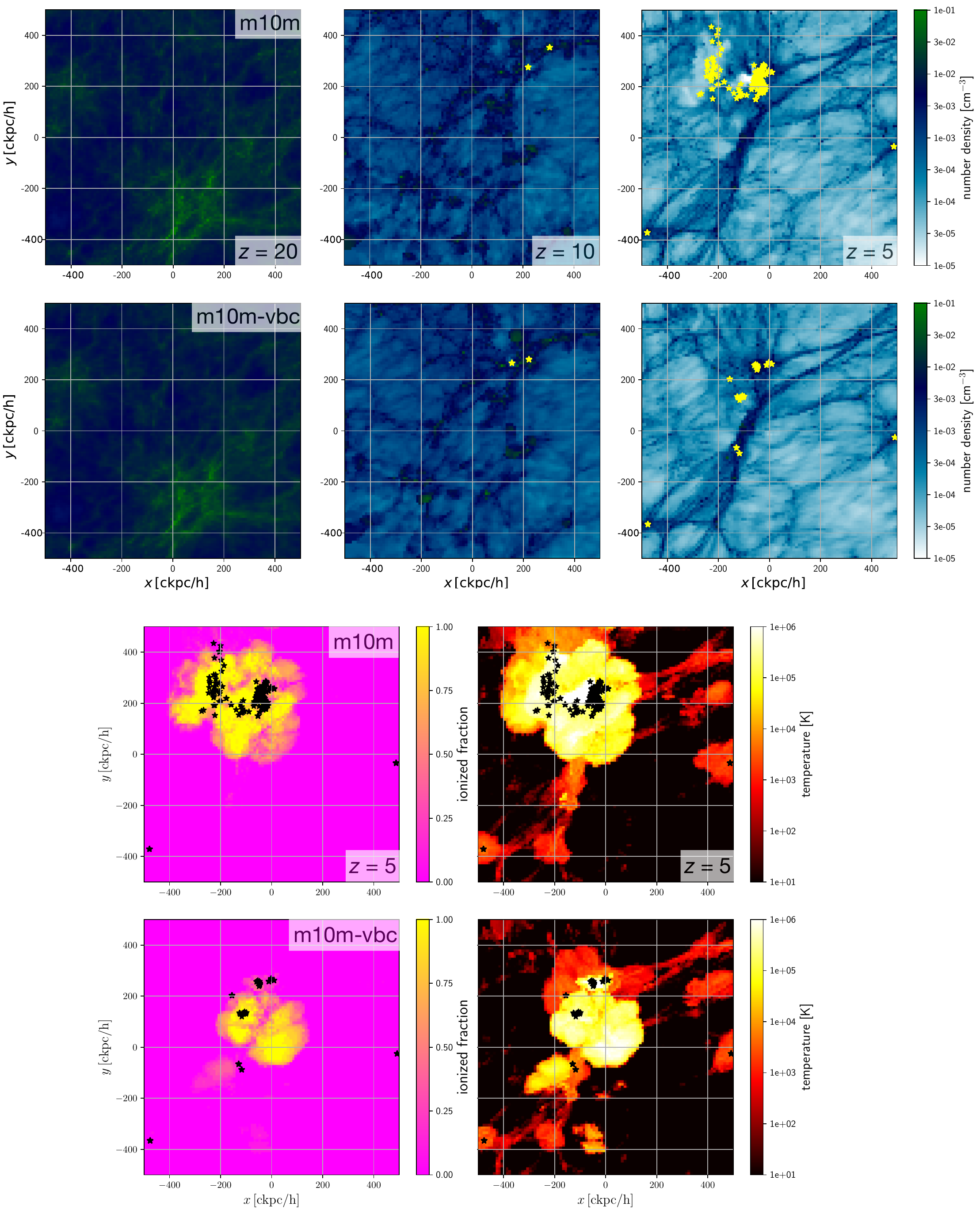}
\caption{Large-scale slice plots (width 100 ckpc/$h$) of the baryon content of simulations m10m (no streaming velocities; first and third row) and m10m-vbc (including streaming velocities; second and forth row). The first two rows show the gas number density, where we can see filaments and halos developing over the redshift range from $z=20$ via $z=10$ to $z=5$. Stars that form in the simulation are marked with filled symbols. In the third and forth row we show the ionization fraction and the temperature of the gas at $z=5$. 
}
\label{fig:slice}
\end{figure*}
We start our analysis by looking at the large-scale structure of our three simulations and their streaming velocity counterparts (labeled with ``vbc''). 
In Figure \ref{fig:slice}, we show slices through the high-resolution regions of simulations m10m (top row) and m10m-vbc (bottom row). 
The baryon density, evolving from redshift $z=20$ to $z=5$ from the first to the third panel, shows similar structure between the streaming and the non-streaming run. The simulation including the streaming velocity exhibits slightly lower density contrasts and generally a more ``washed-out'' behavior than the simulation counterpart without a streaming velocity, although this behavior is not very pronounced. This is consistent with the large-scale behavior of high-redshift simulations that focus on first star formation (see e.g. \citealt{mcquinn12,schauer21}). 

Stars generally form at the highest densities, which lie at the intersection of filaments. The sites for star formation (indicated by star symbols) are similar between the realizations at redshift $z=10$. At redshift $z=5$, star formation is more extended in m10m, leading to a larger ionized region (forth panel) and more high-temperature gas (fifth panel). We investigate the stochasticity of this behavior in the following sections. 
\subsection{Dark matter halos}
\begin{figure}
\centering
\includegraphics[width=1.\columnwidth]{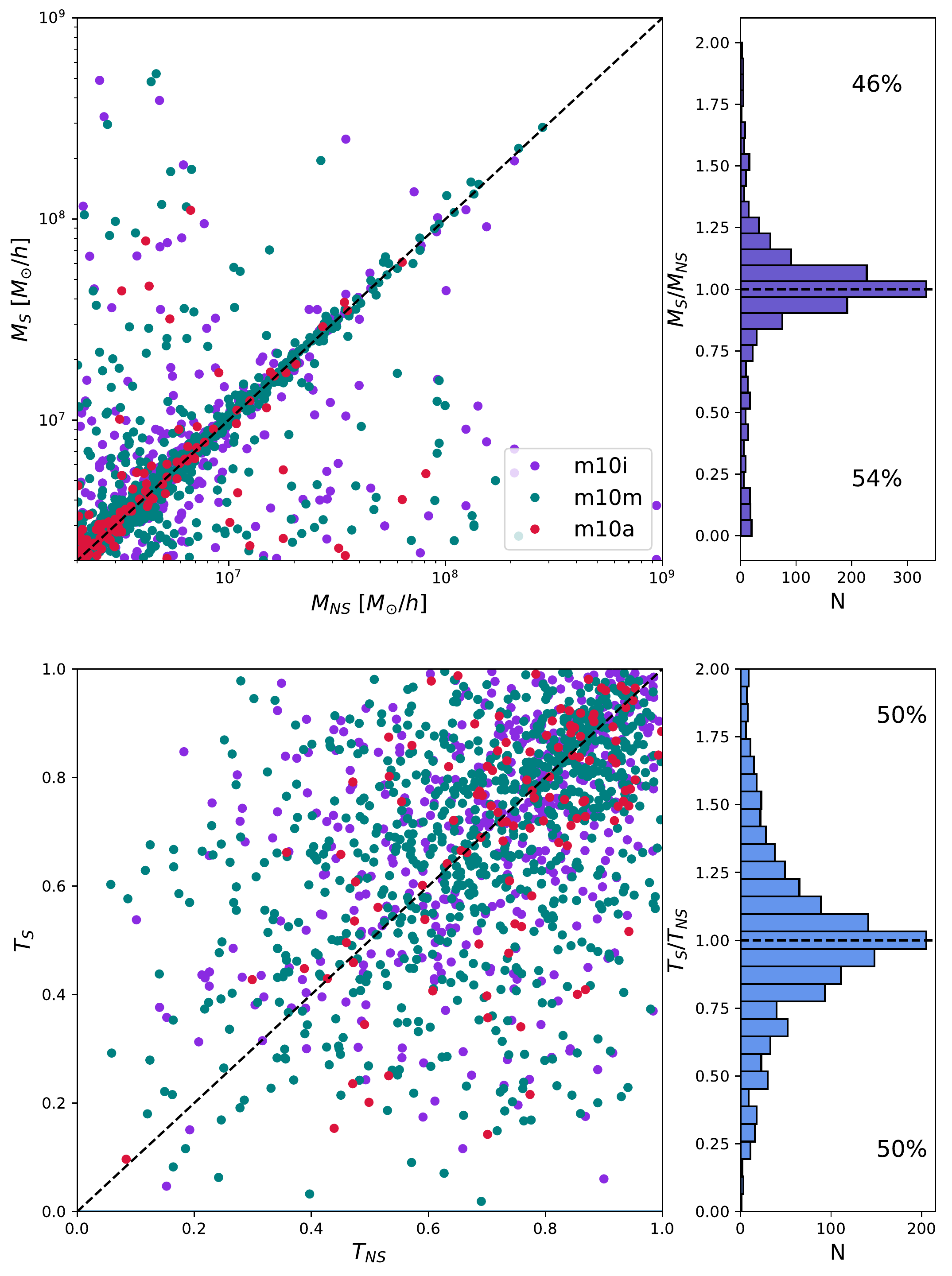}
\caption{Virial masses (top panels) and triaxialities (bottom panels) for the halos in the simulations with streaming velocities (horizontal axis) and their no-streaming velocity counterparts (vertical axis) at redshift $z=5$. The halos follow the 1:1 relation with significant scatter, especially when comparing the triaxialities. 
}
\label{fig:kate}
\end{figure}
To directly compare halos in the streaming velocity simulations and their no-streaming-velocity counterparts, we match halos exceeding $2\times10^6$\,M$_\odot/h$ across paired simulations based on the positions and virial radii of the halos: for each halo in the non-streaming run, we identify its match in the streaming velocity run as the halo that is the closest in terms of distance. To eliminate possible mis-identification of a halo that does \textit{not} have a (spurios) counterpart with a halo far away, we impose the additional criterion that this distance cannot be larger than $\approx 10$~comoving kpc between the candidate match in the streaming versus no-streaming runs. We further restrict the halo pair matches to halos that differ in virial radii by at most 30\% to avoid cross-identifying halos and subhalos. 

Figure \ref{fig:kate} shows two fundamental properties of the halos: their virial mass and their triaxiality $T=(a^2-b^2)/(a^2-c^2)$, where $a \ge b \ge c$ are the primary axes of the halo. The triaxiality determines the shape of the halo from oblate ($T=0$) to prolate ($T=1$). 
We find that streaming and non-streaming halo pairs mostly follow a 
1:1 correlation with some scatter. The scatter is especially high for the triaxiality. These results are in good agreement with e.g. 
\cite{druschke20} at $z=15$, who find that, while shape and mass of the gas is influenced by streaming velocities, the dark matter halo does not change significantly. 

The halo mass function remains almost unchanged between 
the streaming velocity simulations and their non-streaming counterparts (see Panel e in Figures \ref{fig:halo10} and \ref{fig:halo5}). Only a small fraction of the halos are subhalos, the halo mass function is dominated by main halos in both cases. 
\subsection{Gas fraction}
\begin{figure}
\centering
\includegraphics[width=1.\columnwidth]{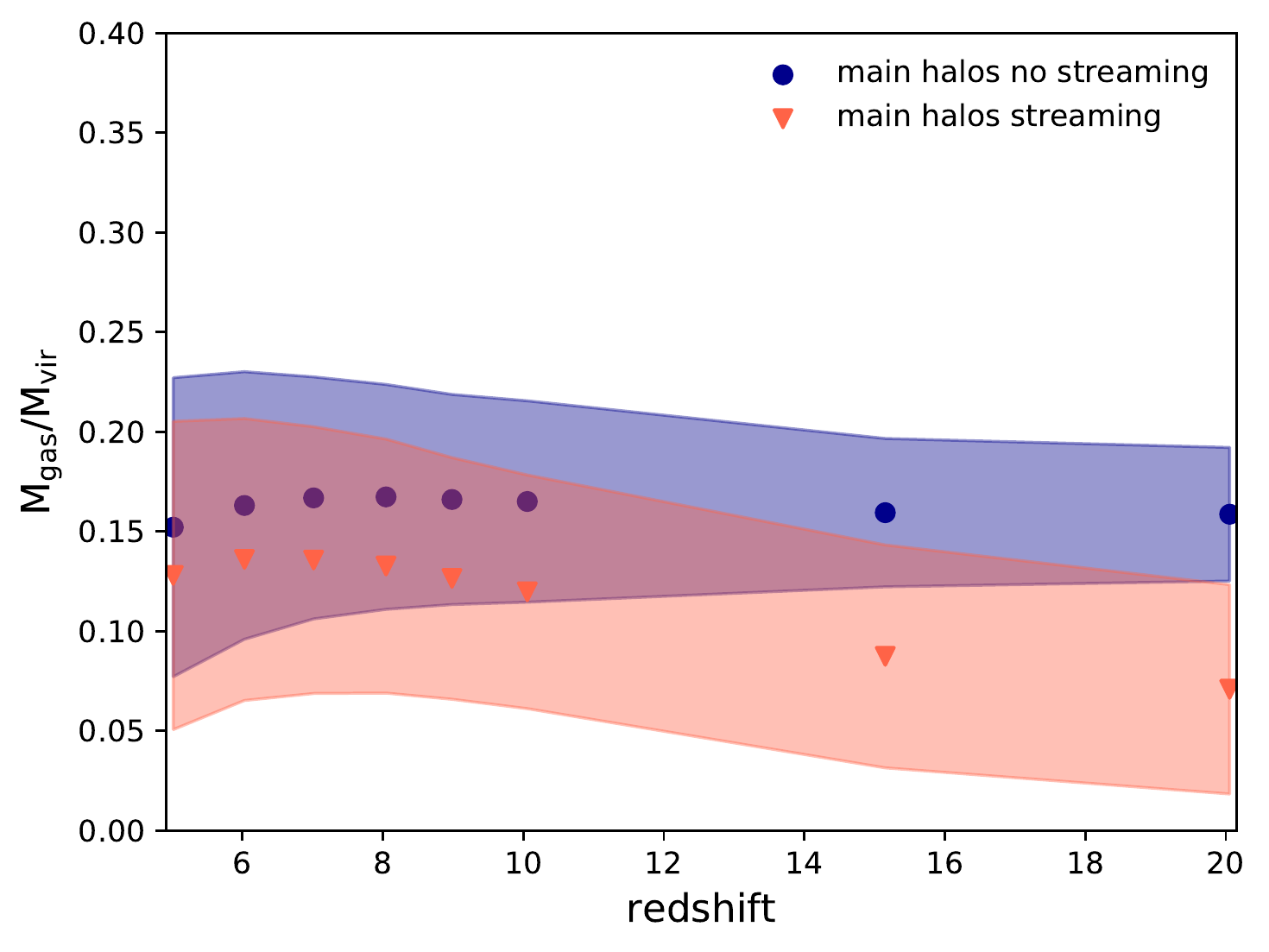}
\caption{Gas fraction as a function of redshift for the halos in streaming velocity regions (orange) and in non-streaming velocity regions (blue). We only show the gas fraction in main halos. Including subhalos leads to much larger 68\% error-bars (shaded regions), with similar mean values (dots/triangles). The gas fraction in streaming velocity halos is smaller than in non-streaming velocity halos, especially at high redshift. }
\label{fig:gasfraction}
\end{figure}

We next turn to the gaseous content of the halos in runs with and without streaming velocities.
We focus on only the main halos in this analysis, applying the general mass threshold of $10^5$\,M$_\odot/h$. 
We see that the gas fraction in the halos in the streaming velocity simulations are strongly reduced compared to the halos of the non-streaming velocity runs (see Figure
\ref{fig:gasfraction}). In the streaming velocity simulations, the mean gas fraction at redshift $z=20$ is only 7\%, significantly below the mean value of the no streaming simulations and the cosmic mean (16\%). In the streaming simulation, the gas fraction rises with decreasing redshift to $\sim$\,13\%, while the non-streaming gas fraction stays relatively constant at around 15\% - 18\%. 
This value is close to the cosmic mean value of $\Omega_{\rm b}/\Omega_{\rm m}=0.16$. We find a large variance, with the gas fraction sometimes exceeding the cosmic mean. To calculate the gas mass of a halo, we take the sum of the mass of the gas particles within the virial radius of the halo, while we take the rockstar-derived virial mass as the halo mass. The dark matter and gas mass within the virial radius differs slightly from the virial mass, introducing an 
bias
in the gas fraction. 

This result is in agreement with the simulations performed by 
\cite{naoz13}, who find a sub-10\% gas fraction in various high resolution runs at redshift $z=20$ and below for a streaming velocity of 1.7~$\sigma_\mathrm{vbc}$. In their highest resolution run with 768$^3$ particles, their baryon fraction rises to $\sim$\,13\% at redshift $z=11$, similar to our result. 
\subsection{Luminous halos}
\begin{figure}
\centering
\includegraphics[width=1.\columnwidth]{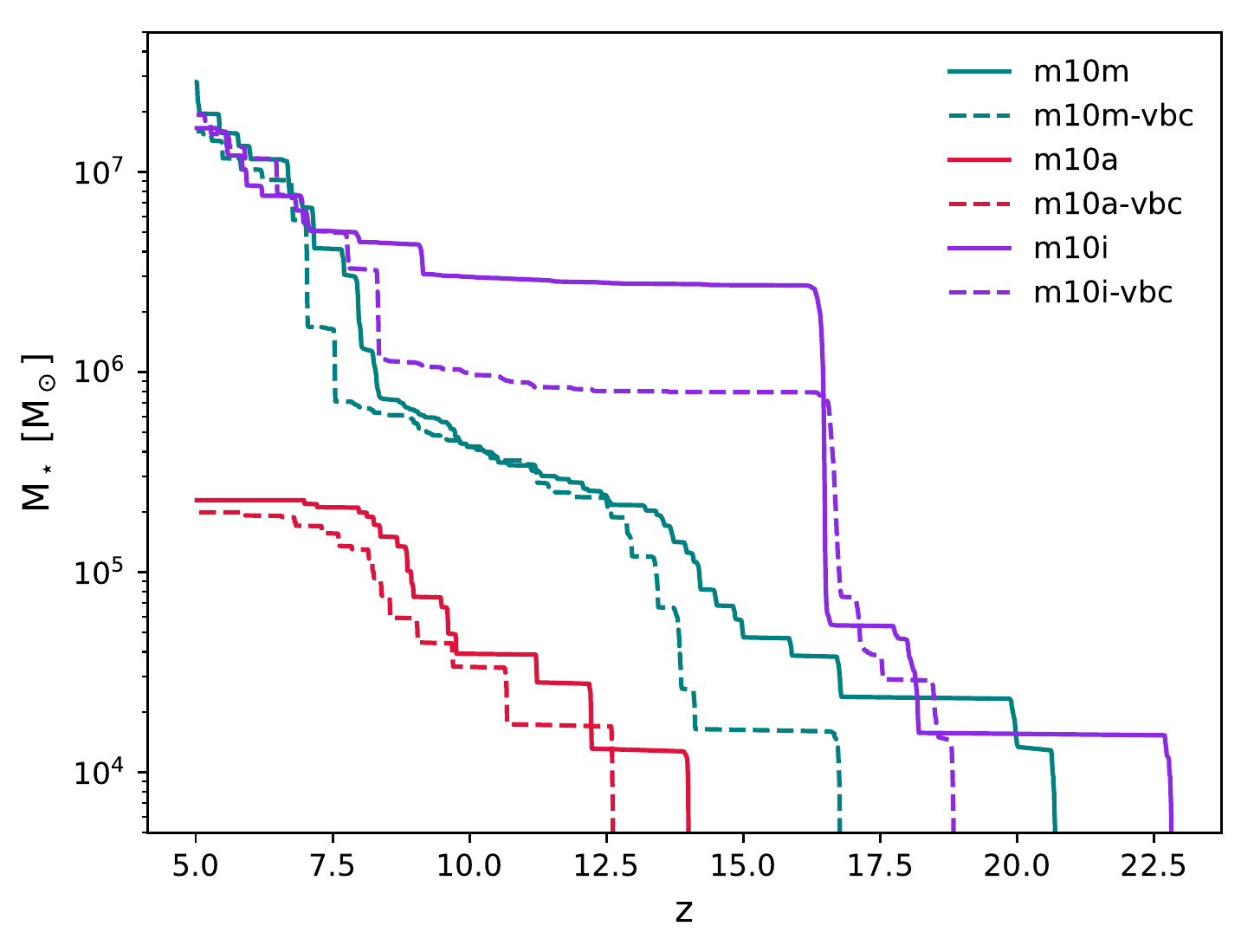}
\caption{Archaeological star formation histories of all six simulations. We show the stellar mass of all stars present at redshift $z=5$ as a function of their formation redshift. Stars formed in streaming velocity simulations are shown with dashed lines, stars formed in no streaming velocity regions are shown with solid lines. While star formation in streaming velocity simulations is delayed and lacks behind the non-streaming velocity stellar mass counterpart, it is able to catch up to a similar level at redshift $z=5$.}
\label{fig:sfh}
\end{figure}
Globally, star formation is delayed in streaming velocity regions, compared to regions of the Universe without a streaming velocity. This can be seen in Figure \ref{fig:sfh}, where we show the cumulative stellar mass at redshift $z=5$. 
In all three sets of simulations, the formation of the first star is delayed by approximately 50\,Myr (from $z=14$ in m10a to $z=12.6$ in m10a-vbc, from $z=22.9$ in m10i to $z=18.9$ in m10i-vbc, and from $z=20.7$ in m10m to $z=16.8$ in m10m-vbc).  
In this initial starburst, tens of thousands of solar masses of stars form. In simulations m10i-vbc and m10a-vbc, the initial starburst is more massive than in the no streaming velocity counterpart simulations, and the total stellar mass in the streaming velocity simulations exceeds the stellar mass in the non-streaming velocity simulation for 10-20\,Myr. Generally, the simulations without a streaming velocity have a higher stellar mass by a factor of a few down to redshift $z\approx 8$, when this difference becomes small and merely stochastic. This diminishing difference implies that the star formation rate in the streaming runs is comparable to or greater than that of the non-streaming runs after the initial onset of star formation, though this effect is relatively modest.

\begin{figure}
\centering`
\includegraphics[width=1.\columnwidth]{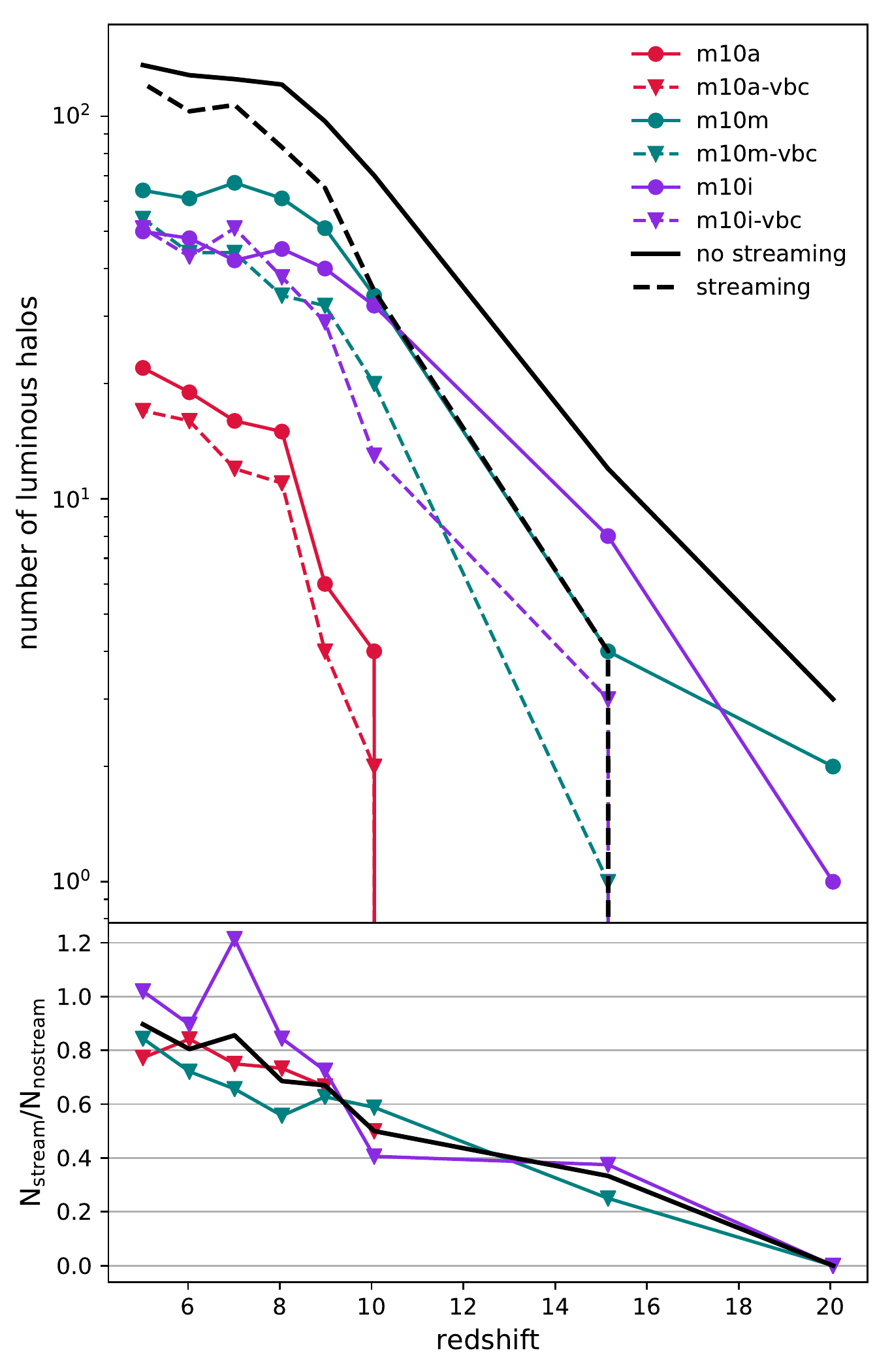}
\caption{\textit{Top:} Number of luminous halos as a function of redshift for streaming velocity simulations (dashed lines) and no streaming velocity simulations (solid lines). In all individual simulations (thin lines), the number of luminous sources is higher for the no streaming velocity regions, resulting in the total number of sources (think lines) also being higher. \textit{Bottom:} Fraction of luminous halos in streaming to those in non-streaming velocity simulations. The fraction increases from zero to 90\% over the redshift range from $z=20$ to $z=5$. }
\label{fig:nlum}
\end{figure}

At the same time, the number of luminous halos increases with time. The global delay of star formation in the streaming velocity simulations leads to a global delay in the emergence of galaxies. 
In Figure \ref{fig:nlum}, we show the number of halos that host stars as a function of redshift for the  six individual simulations and the streaming and no streaming simulations combined. At the highest redshift, $z=20$, only simulations m10m and m10i have two and one halo(s) with stars, respectively, while no halo in a streaming velocity simulation has formed stars. At about redshift $z=10$, there are half as many galaxies in the streaming velocity simulations as in the no streaming velocity simulations. This fraction increases until it reaches 90\% at redshift $z=5$.

\begin{figure*}
\centering
\includegraphics[width=1.\columnwidth]{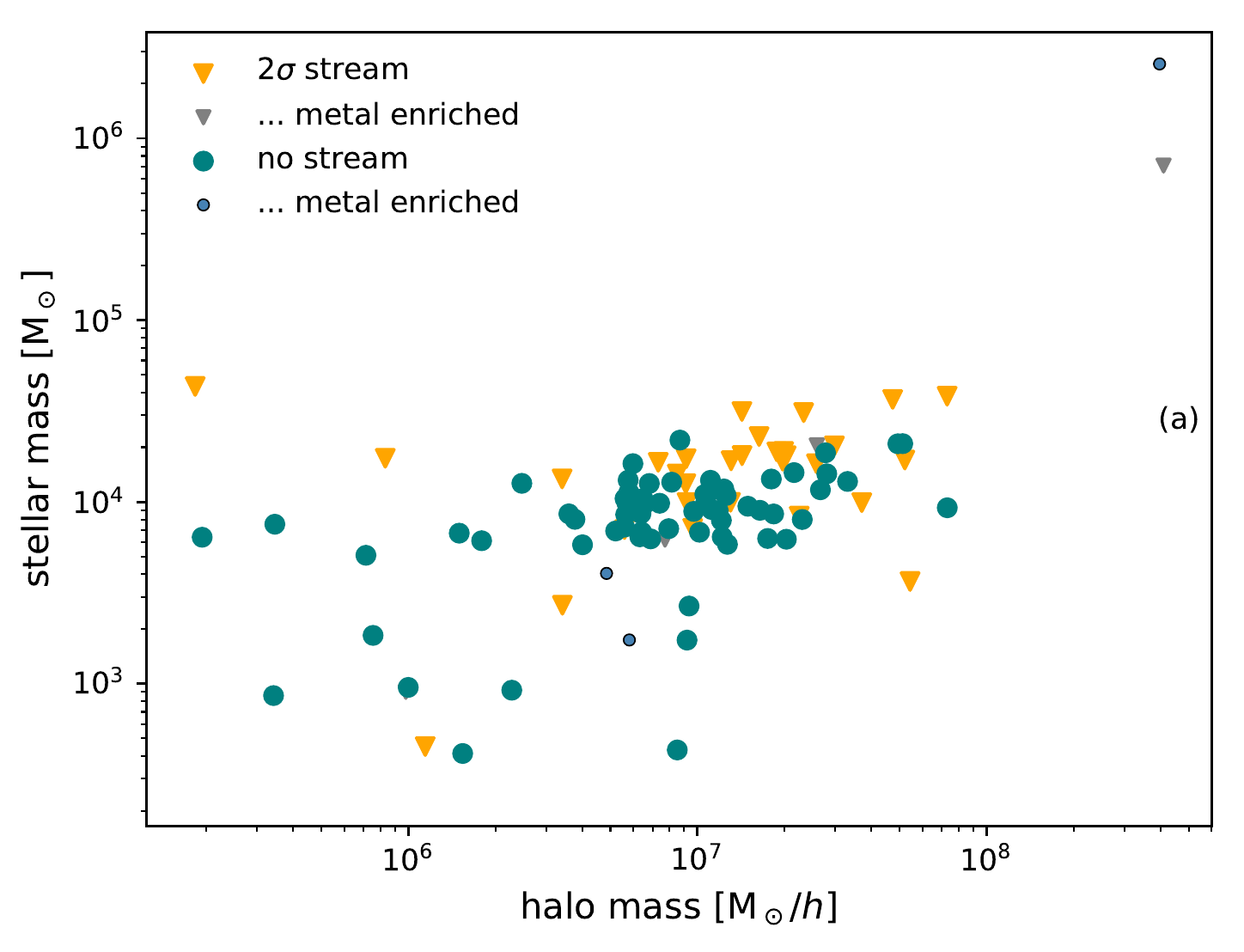}
\includegraphics[width=1.\columnwidth]{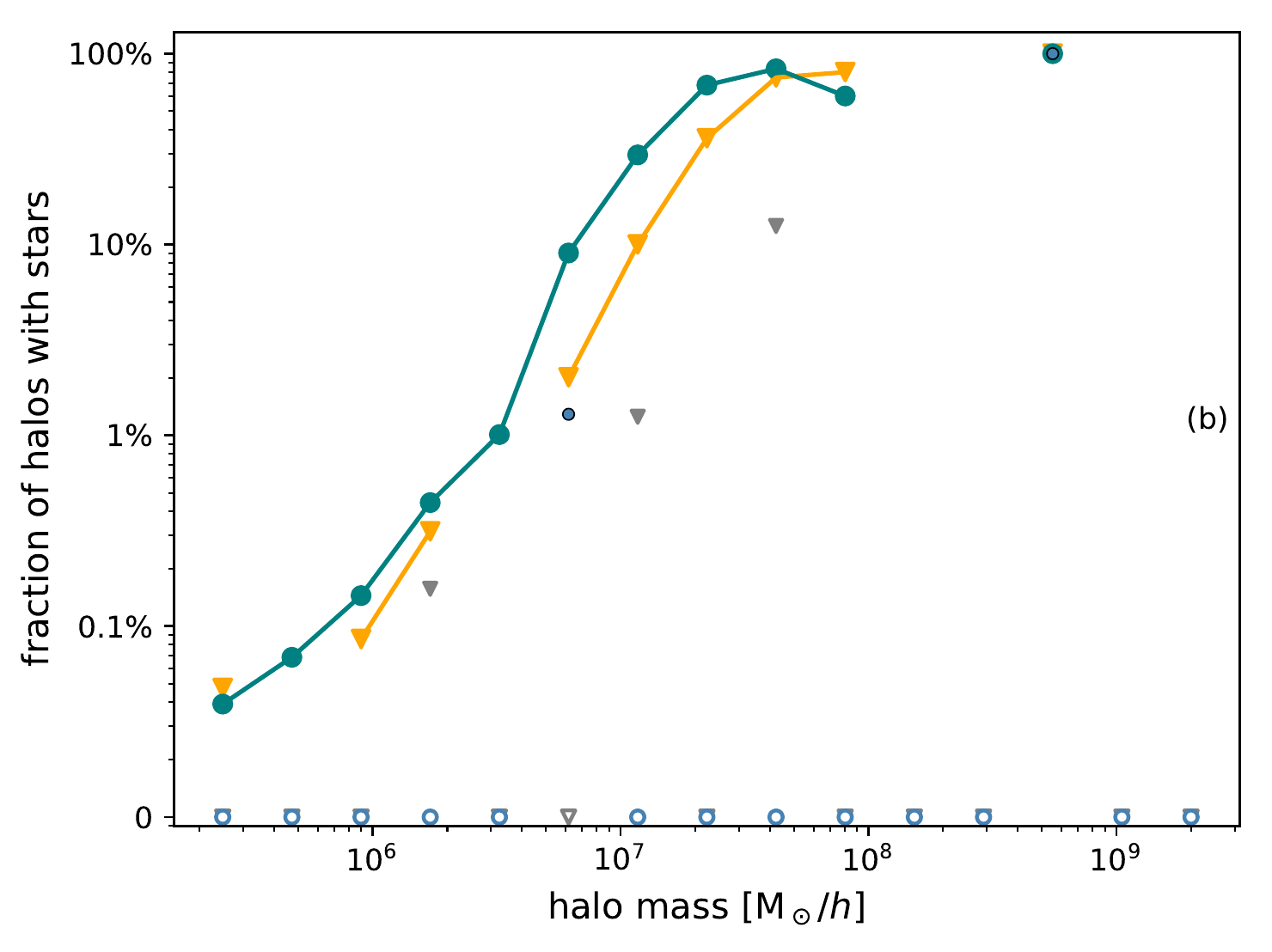}\\
\includegraphics[width=1.\columnwidth]{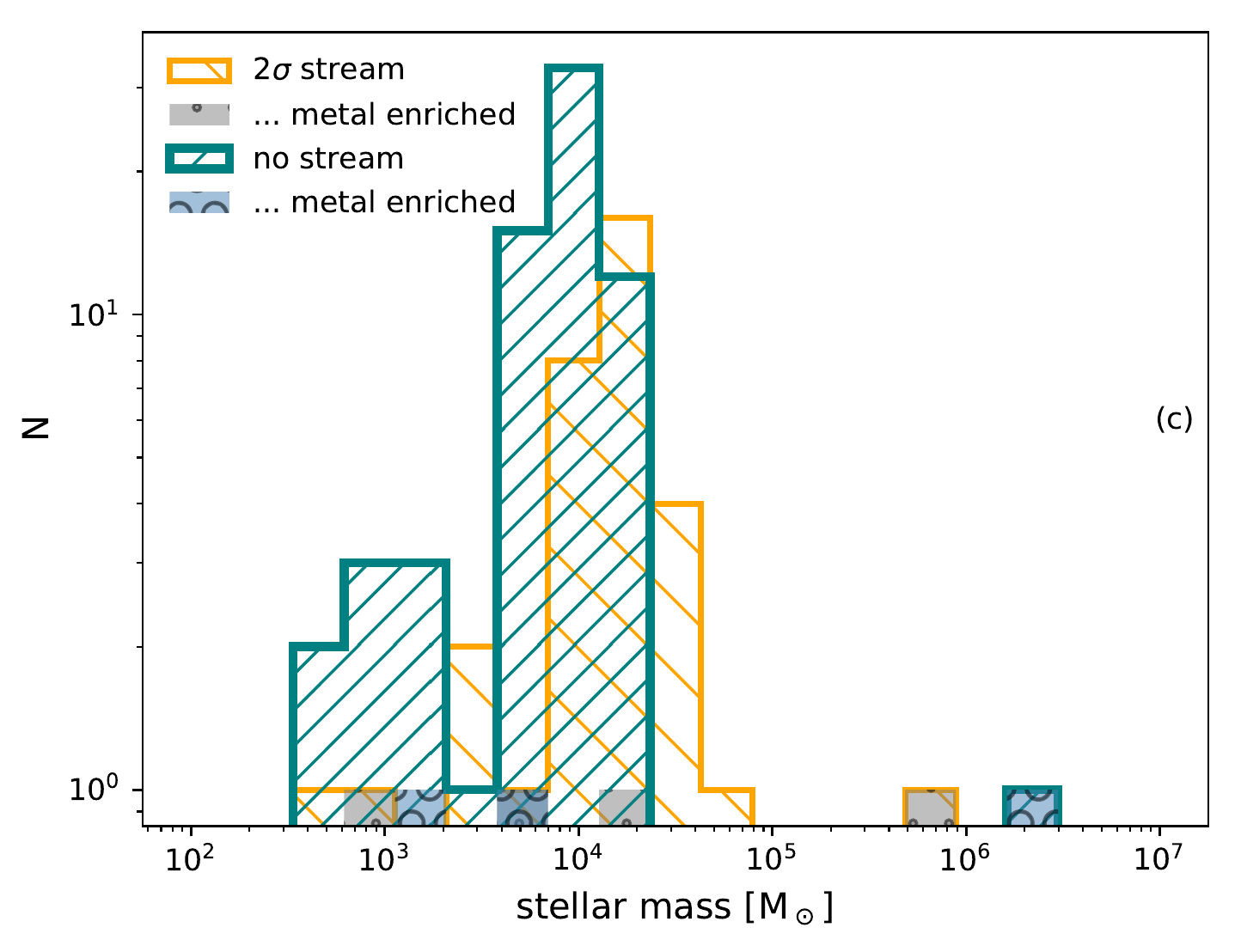}
\includegraphics[width=1.\columnwidth]{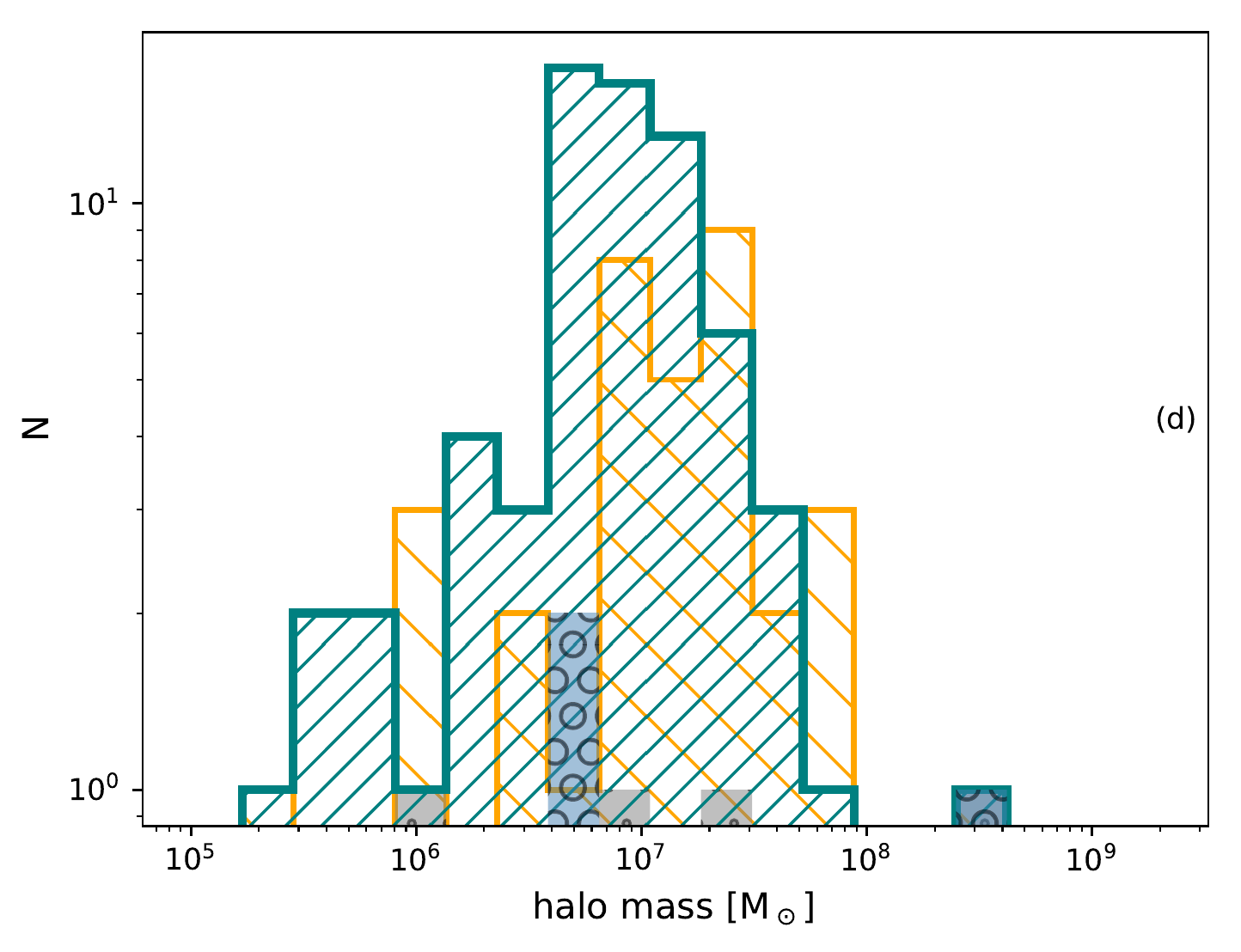}\\
\includegraphics[width=1.\columnwidth]{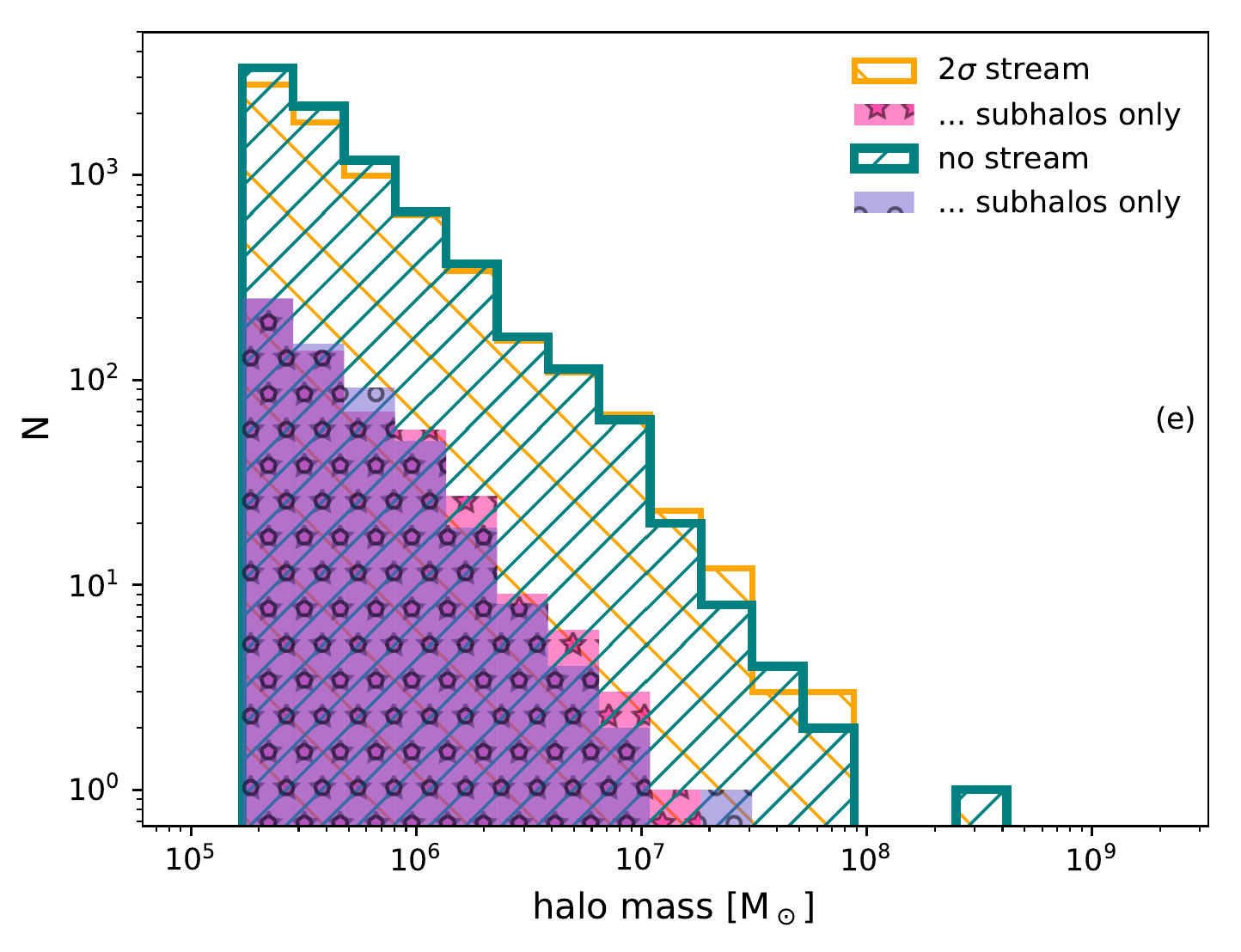}
\includegraphics[width=1.\columnwidth]{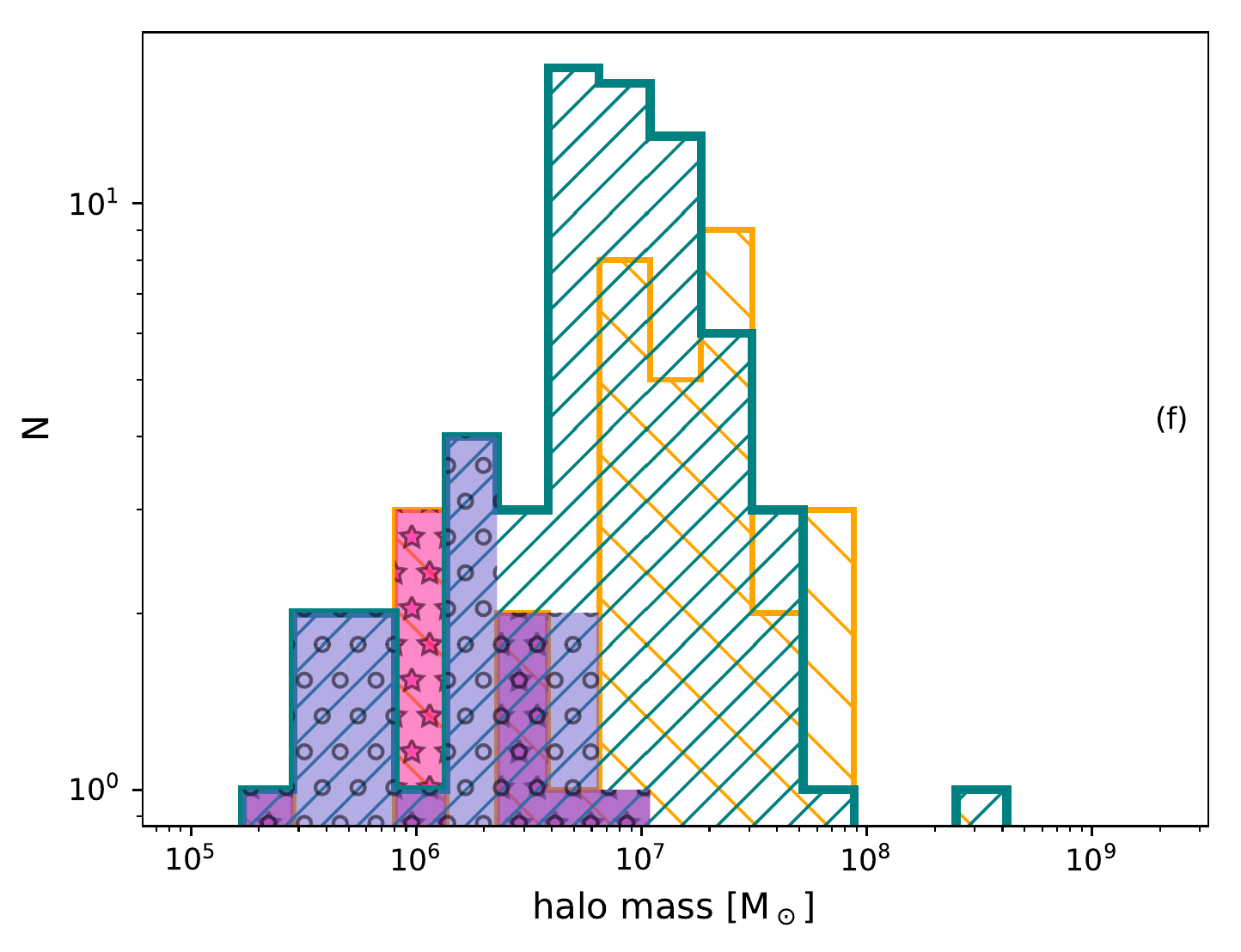}
\caption{Various halo statistics for redshift $z=10$. From top left (Panel a) to bottom right (Panel f): (a) stellar mass-halo mass relation, (b) fraction of halos that host stars as a function of halo mass, (c) histogram of stellar masses in halos, (d) histogram of halo masses of halos that host galaxies, (e) histogram of all dark matter halos, (f) histogram of halo masses of halos that host galaxies. Halos from no streaming velocity simulations are shown with teal (points), halos from streaming velocity simulations are shown with yellow (triangles). Panels (a) - (d) distinguish between metal free (light color) and metal enriched (gray / black component) stars. Panels (e) and (f) distinguish between all halos and subhalos (pink / purple). 
}
\label{fig:halo10}
\end{figure*}
\begin{figure*}
\centering
\includegraphics[width=1.\columnwidth]{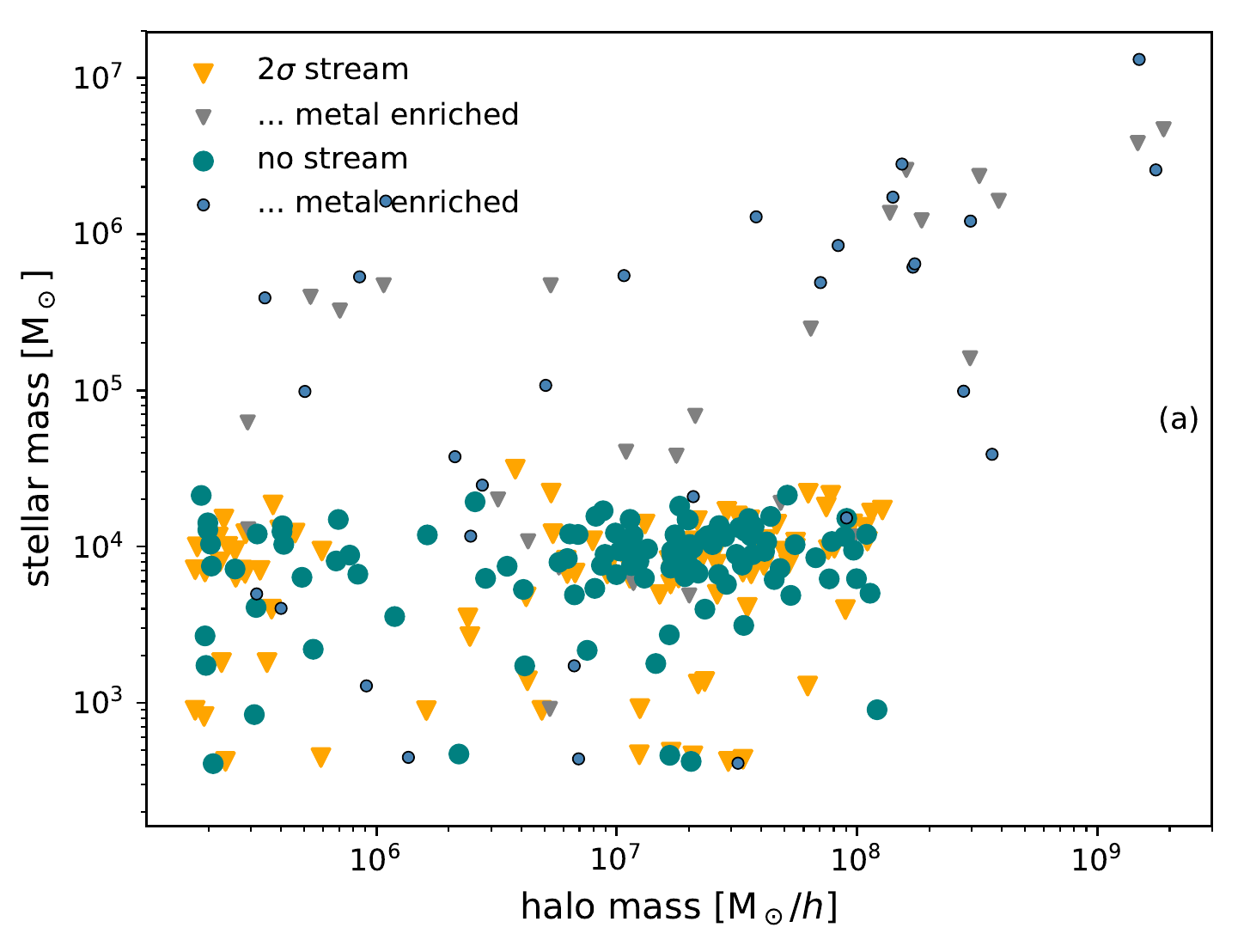}
\includegraphics[width=1.\columnwidth]{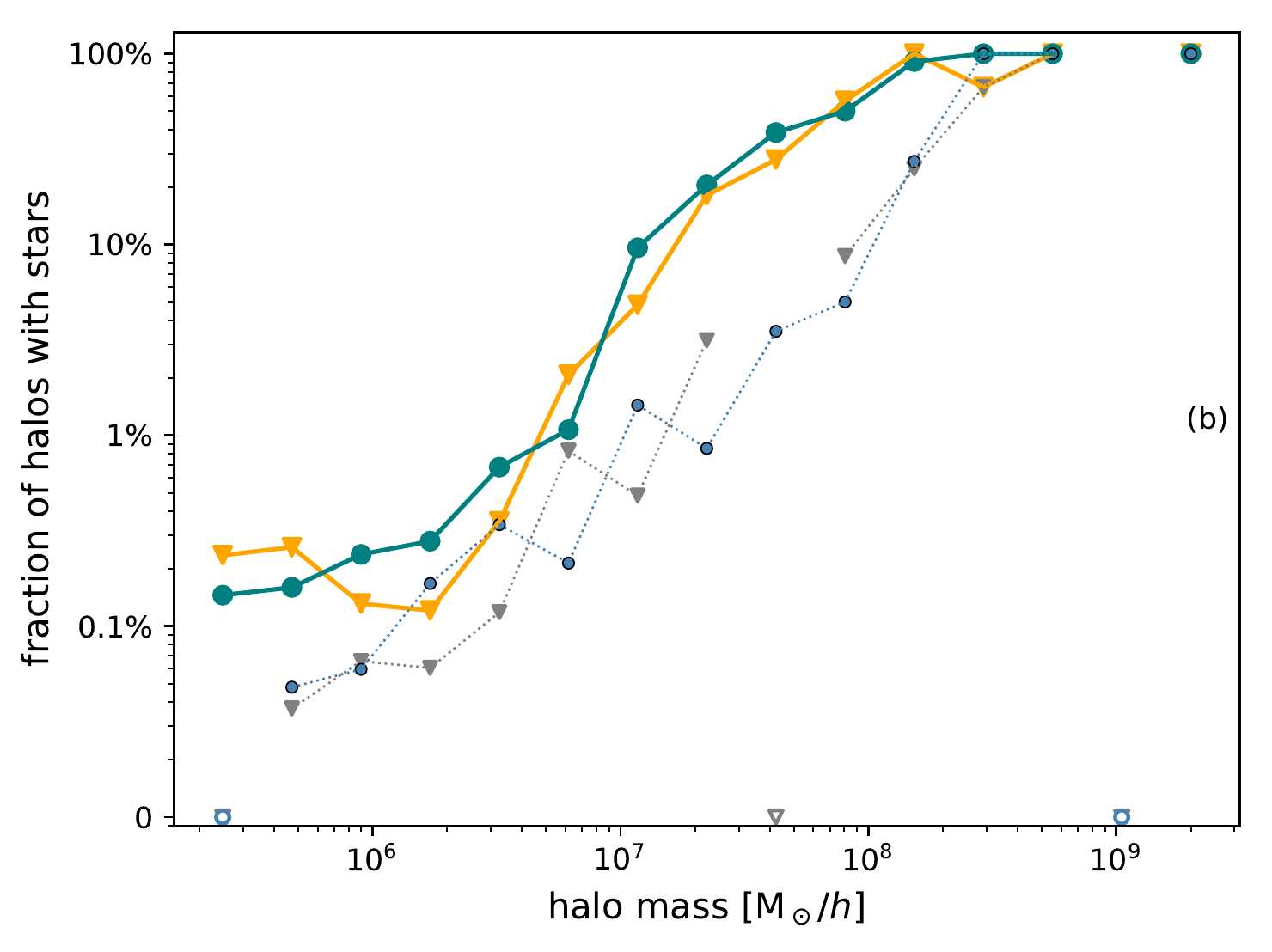}\\
\includegraphics[width=1.\columnwidth]{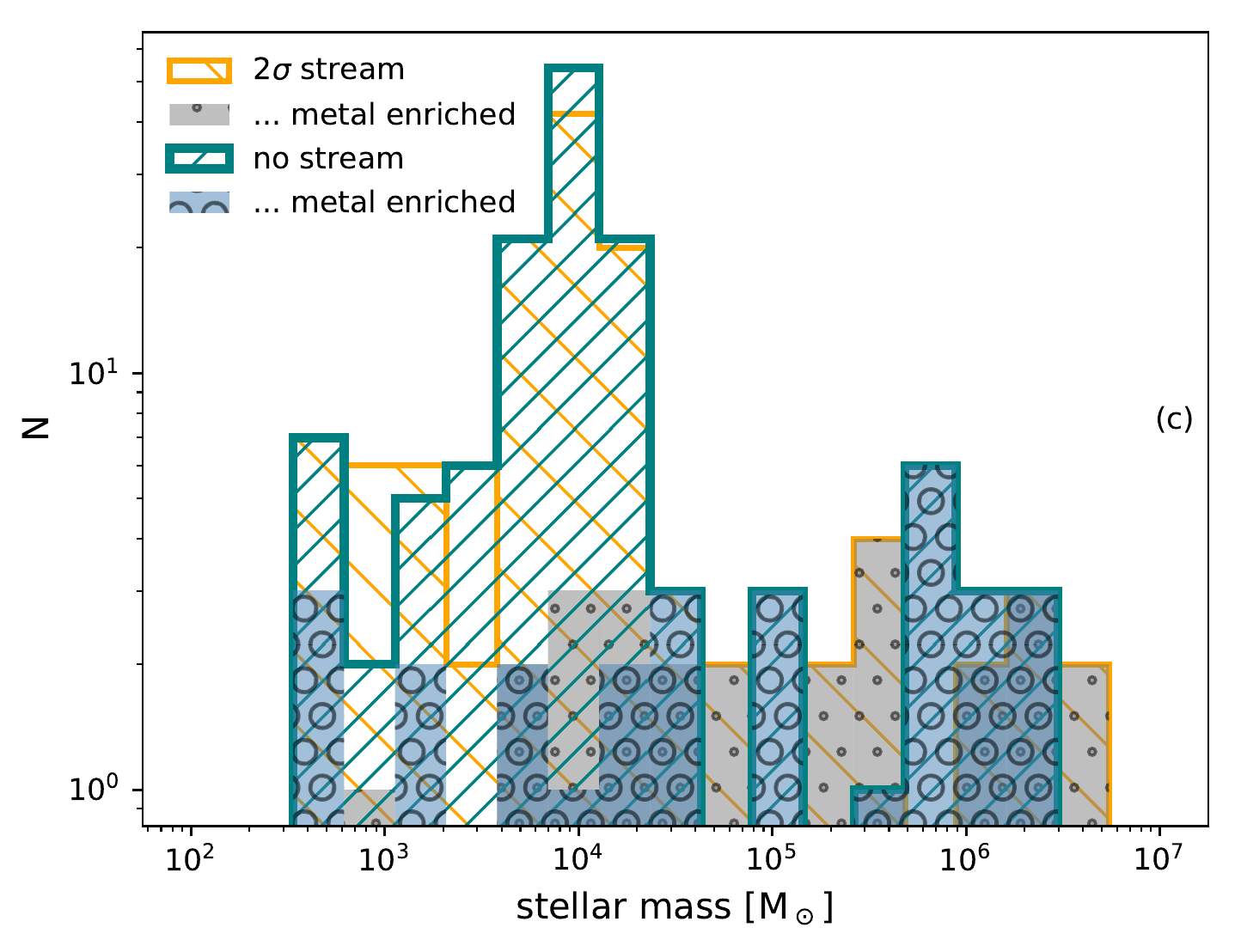}
\includegraphics[width=1.\columnwidth]{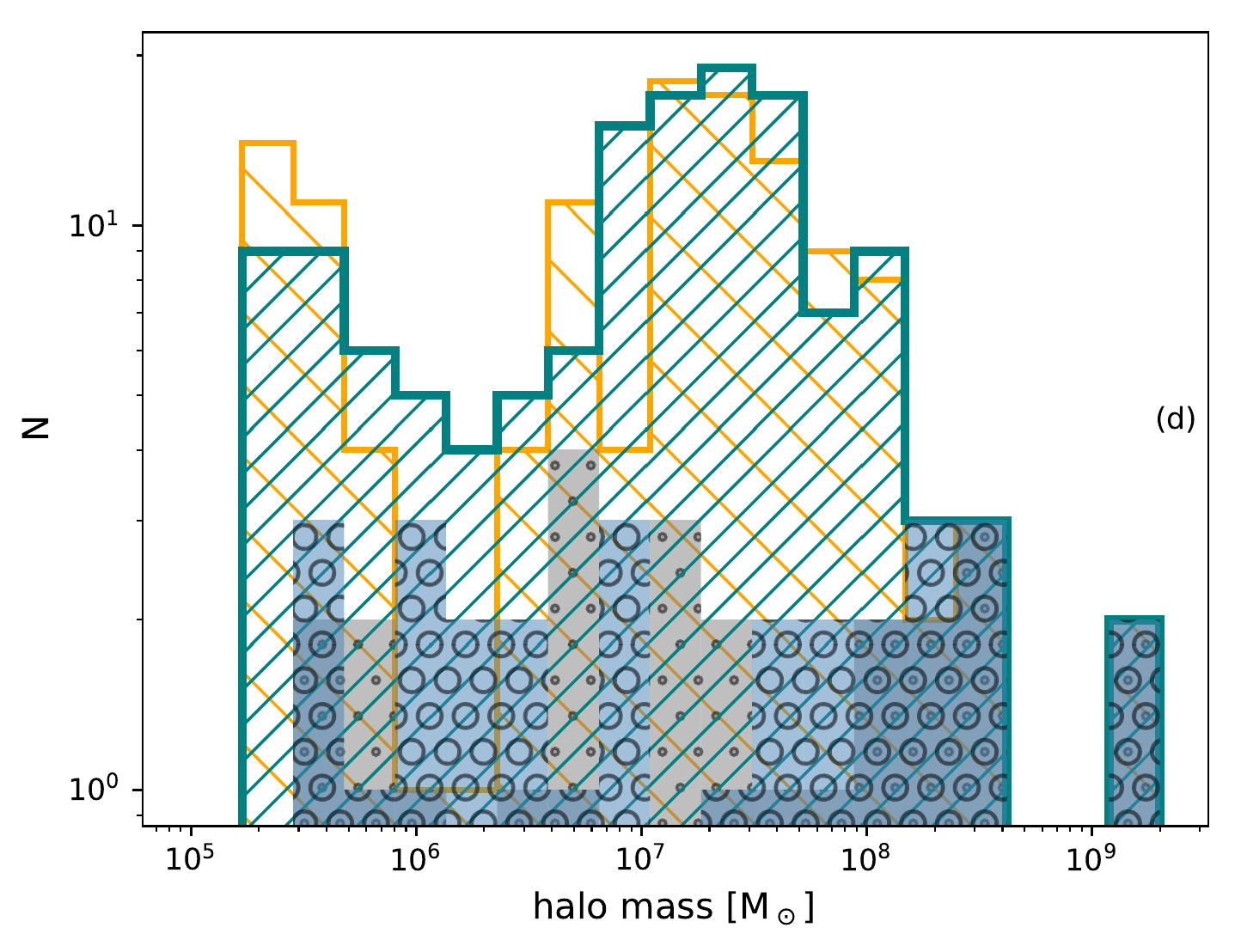}\\
\includegraphics[width=1.\columnwidth]{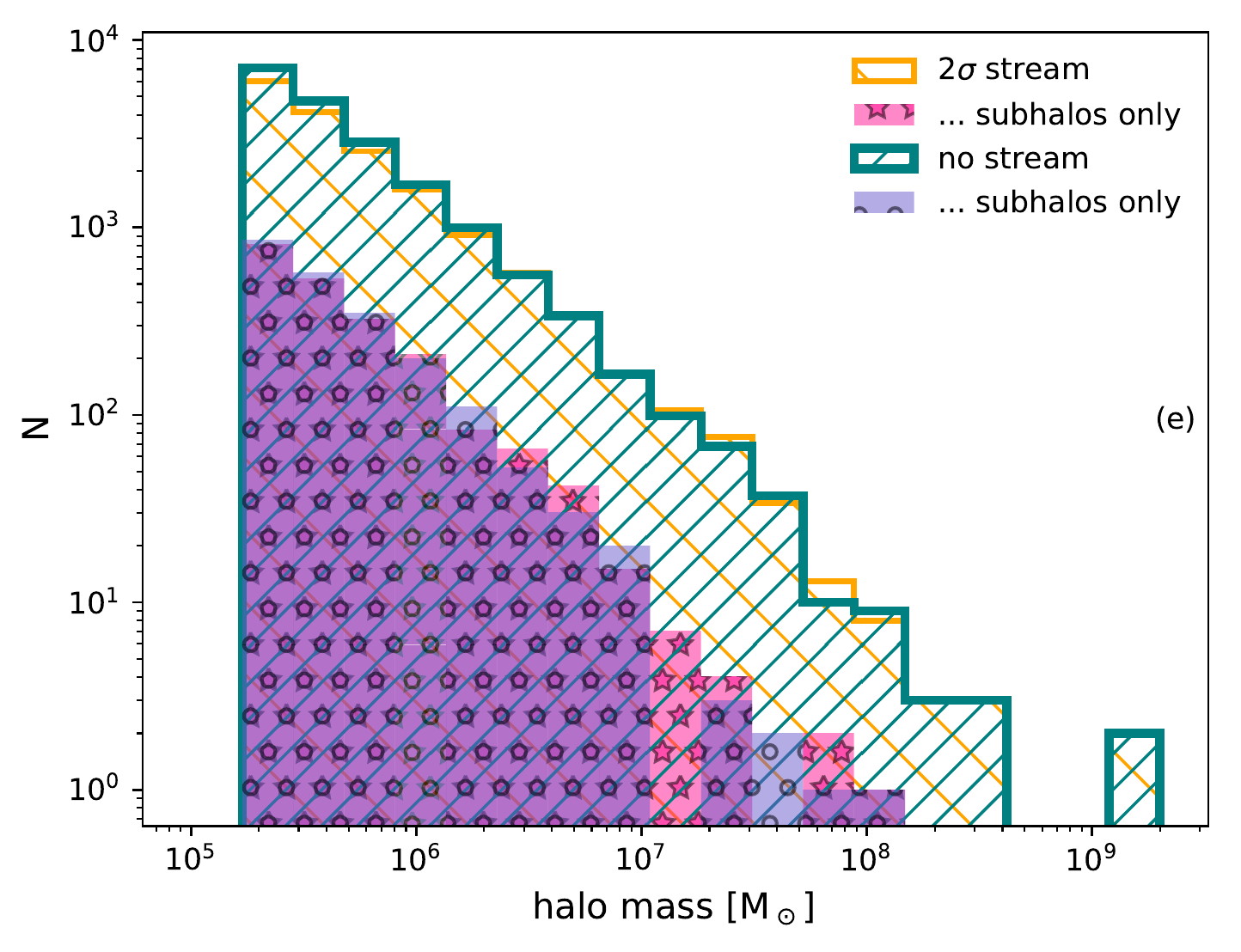}
\includegraphics[width=1.\columnwidth]{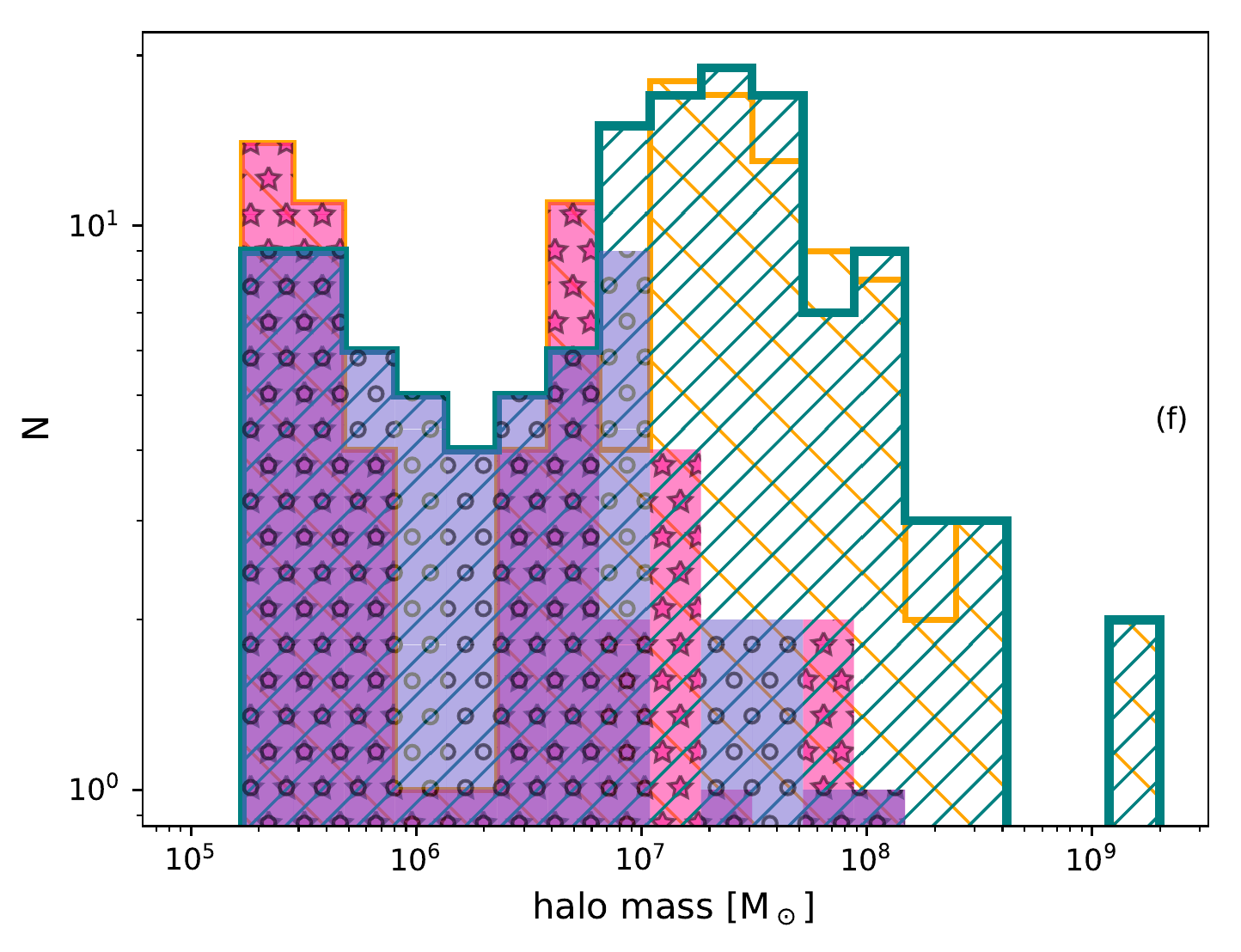}
\caption{Halo statistics, same as in Figure \ref{fig:halo10}, for redshift $z=5$.}
\label{fig:halo5}
\end{figure*}

In a next step, we focus on the statistics of the halos at redshifts $z=10$ and $z=5$ in Figures~\ref{fig:halo10} and \ref{fig:halo5}. 
Most halos form stars during an initial star burst with a total mass of $10^4$\,M$_\odot$. At redshift $z=10$ the stellar mass per halo is higher in streaming velocity simulations (panels a and c of Figure \ref{fig:halo10}), with little correlation to the halo mass. This means that the few halos in a streaming velocity simulation which are able to form stars have an initially larger starburst. At later times, at redshift $z=5$, the metal free population is similar in both streaming velocity and no streaming velocity runs, with the probability distribution again peaking around $10^4$\,M$_\odot$ (panels a and c of Figure \ref{fig:halo5}). 

Only a small fraction of halos is metal enriched, down to redshift $z=5$ (compare panels b of Figures \ref{fig:halo5} and \ref{fig:halo10}).
This, however, is likely a consequence of not explicitly modeling metal enrichment from Pop~III stars, with a top-heavy IMF \citep[e.g.,][]{Jeon2015}.
At our lowest redshift, we see a weak correlation between halo mass and stellar mass, where more massive halos host larger stellar masses. 

Above a halo mass of $10^8$\,M$_\odot$, almost all halos have formed stars. This fraction decreases rapidly with decreasing halo mass. In an intermediate halo mass range of $10^6$\,M$_\odot$ - $10^8$\,M$_\odot$, the fraction of no-streaming halos hosting stars is higher than the fraction of streaming halos hosting stars, with the behavior more visible at earlier times. At small halo masses of a few $10^5$\,M$_\odot$, the fraction of luminous halos decreases to below (redshift $z=10$) 0.1\%/ slightly above (redshift $z=5$) 0.1\%.

Both mean and median mass of star forming halos are higher in case of the halos being embedded in the streaming velocity simulation - by about a factor of 2 at redshift $z=10$, 
and by only maximally 10\% at redshift $z=5$. The same is true when focusing on the star forming main halos only, with the mean and median being of up to a factor 2 larger at redshift $z=10$ and of only up to 15\% larger at redshift $z=5$. 
These values are smaller than the factor of 6 increase reported by \cite{schauer21}, averaged over a redshift range of $z=22$ through $z=14$, and also smaller than the factor of three increase in halo mass reported by the early study by \cite{greif11} in the redshift range of $z=30-15$. 

\begin{figure*}
\centering
\includegraphics[width=1.\columnwidth]{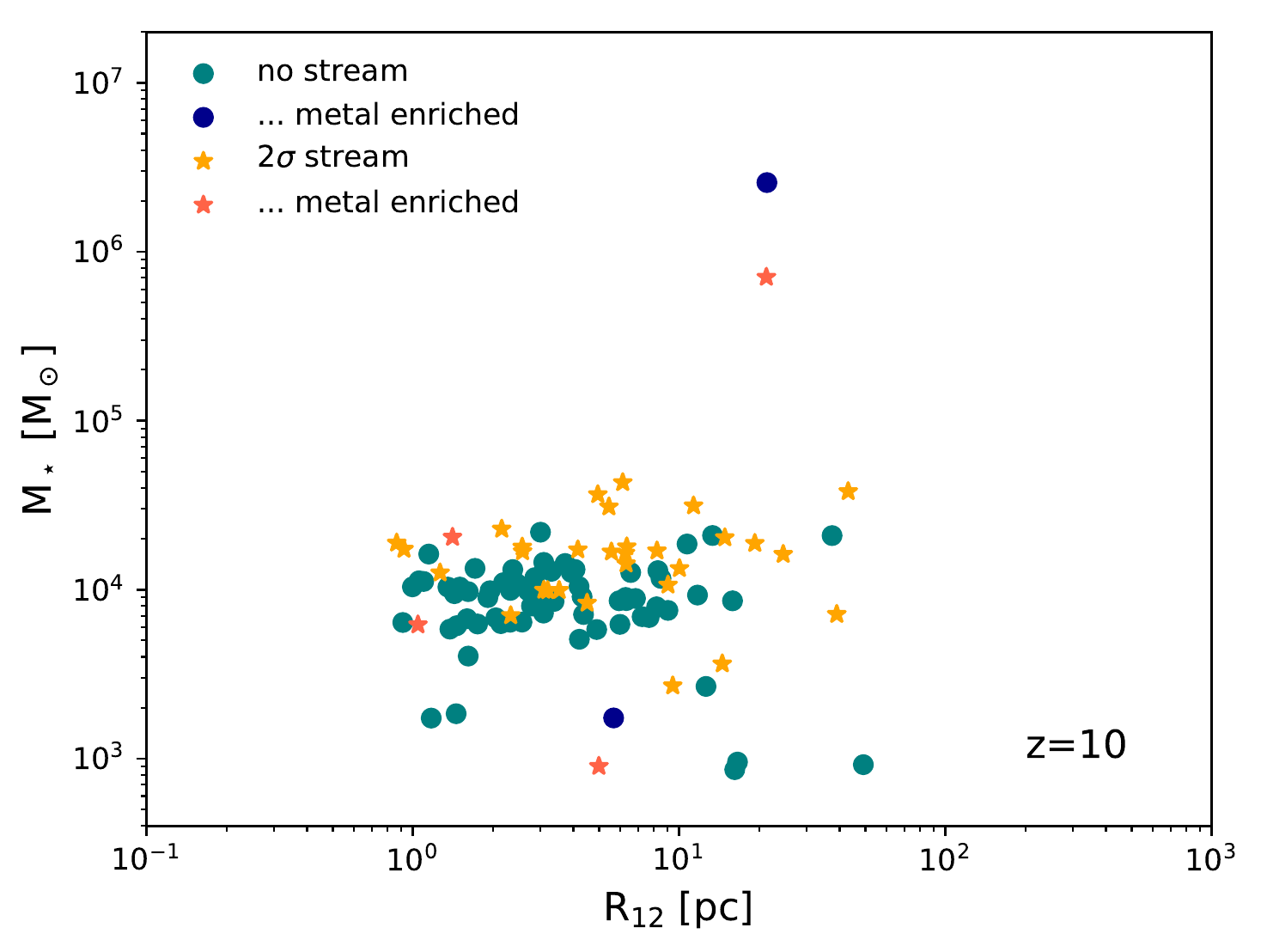}
\includegraphics[width=1.\columnwidth]{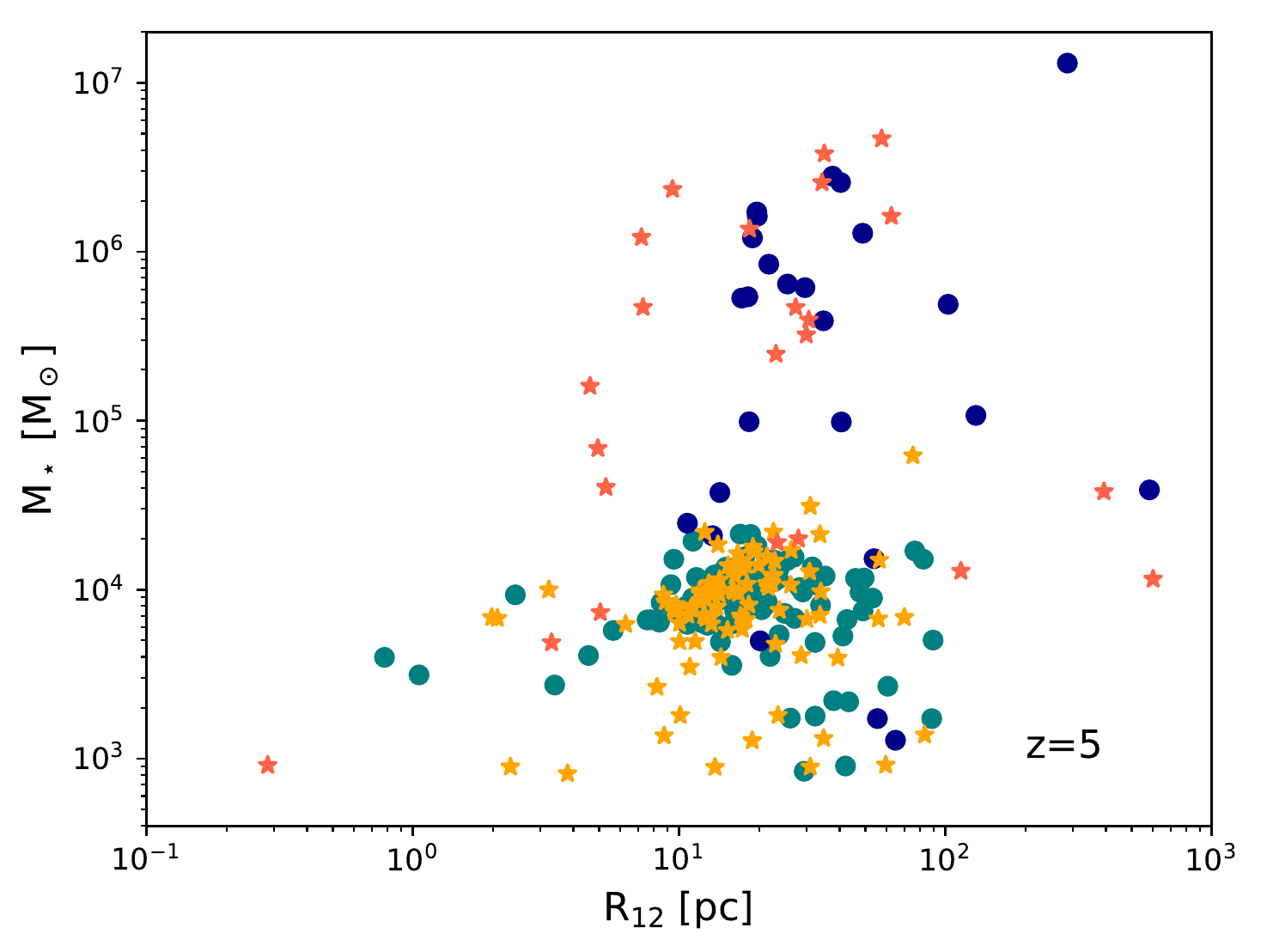}
\caption{Stellar mass as a function of the half-mass radius of the stars. Halos in no streaming velocity regions are shown with blue/teal dots, halos in streaming velocity regions are shown with yellow/orange stars. While the half-mass radii range over several orders of magnitudes, we do not see significant differences between streaming / nonstreaming halos at neither redshift $z=10$ (left panel) nor redshift $z=5$ (right panel). Metal enriched galaxies have on average larger half-mass radii, as multiple star formation events spread the stars over larger radii.}
\label{fig:r12}
\end{figure*}
The size of the star forming regions spans several orders of magnitude. We calculate the half-mass radius of stars associated with a halo (see Figure \ref{fig:r12}) and find them ranging between 1\,pc and 60\,pc at redshift $z=10$. 
At redshift $z=5$, this spread is even larger, from sub-parsec scales to almost 1\,kpc. The very small half-mass radii are associated with a small stellar population of around 1000\,M$_\odot$, below the mass of a globular cluster. We see little difference in the streaming and the non-streaming half-mass radii, at both redshifts.

\section{Summary and Conclusions}
\label{sec:conclusions}
In our study, we  have examined the effects of streaming velocity on a set of three dwarf galaxy progenitors from the FIRE-3 
simulation suite -- studying several thousand progenitor halos -- between the redshift of first star formation down to the epoch of reionization. 
In agreement with multiple high-redshift studies, we find that the onset of star formation is delayed, by approximately 50\,Myr in all three simulation pairs. 
The total mass in stars is subsequently smaller down to a redshift of $z\sim8$. The number of luminous sources (galaxies inside dark matter halos) in streaming velocity regions vs. non-streaming velocity regions increases from a few percent at redshift $z=20$ to 50\% at $z=10$, until it almost becomes equal at the lowest redshift probed here ($7>z>5$).

At these low redshifts, we also see no significant change in the minimum halo mass for star formation. The reason for this is only partially due to metal enrichment, galaxy formation at lower redshift is increasingly unaffected by the decreasing streaming velocity. We conclude that while streaming velocities significantly affect the high-redshift Universe during the Dark Ages, the effects become less important even in small galaxies at the epoch of reionization. Quantities such as the minimum halo mass for star formation, the half-mass radius of the galaxy or the stellar mass content show very little difference in a region of high-streaming velocity as compared to a region of no streaming velocity at redshift $z=5$. 
This is consistent with the results from \cite{gutcke22}, who do not include streaming velocities explicitly in their dwarf galaxy simulation,  but instead alter the halo mass threshold for Pop~III star formation - one of the high-redshift effects of streaming velocities. While the onset of first star formation is delayed for a higher halo mass (comparable to the simulations with a streaming velocity), the final stellar and halo masses are not significantly altered. 

Our simulations do not self-consistently follow reionization but instead include a global reionization background. This leads to the simulation volume not yet being reionized by redshift $5$. Our results are therefore limited to pre-reionization predictions \citep[see, e.g.,][for assessing this caveat]{Milosavljevic2014}.

Even though the effects of streaming velocities on properties such as the luminosity function get smaller with redshift, we cannot exclude for certain the possibility of persistent observable features. 
We run our simulations with standard FIRE-3 physics which does not follow Pop~III star formation explicitly with a network of primordial chemistry and an altered initial mass function for metal free stars \citep[see, e.g.,][]{jaacks19}. 
Different chemical abundance patterns, for example different levels of r-process enrichment \citep{jeon21}, could be found if distinguishing between Pop~III and Pop~II star formation and nucleosynthesis \citep{jeon17}, possibly allowing us to determine if a first galaxy was born in a region of high or low streaming velocity. 

In the future, we plan on including the formation of Pop~III stars in the simulation, as carried out in a recent study by \citet{sanati22} with the SPH code GEAR. 
It will be interesting to see if this result is influenced by pure Pop~III star formation, since Pop~III stars are expected to survive in pristine clumps down to redshift $z=6$ \citep{liu20}. 

On the observational side, this era of the very first onset of star formation is out of reach  for current and upcoming facilities \citep{jeon19}. A possible exception could be extreme gravitational-lensing events, such as the Earendel stellar source \citep[][]{welch22,schauer22earendel}, or a potential Pop~III star cluster \citep{vanzella20}. These single objects, however, fail to characterize the full population of minihalos and small first galaxies. Direct observations of this ultimate high-redshift frontier of star and galaxy formation calls for 100\,m-sized 
extraterrestrial telescopes \citep{angel08,rhodes20, schauer20a}.

\section*{Acknowledgments}
We would like to thank Aaron Smith 
for helpful comments. MBK acknowledges support from NSF CAREER award AST-1752913, NSF grants AST-1910346 and AST-2108962, NASA grant 80NSSC22K0827, and HST-AR-15809, HST-GO-15658, HST-GO-15901, HST-GO-15902, HST-AR-16159, and HST-GO-16226 from the Space Telescope Science Institute, which is operated by AURA, Inc., under NASA contract NAS5-26555. JSB was supported by NSF grant AST-1910346.
AW received support from: NSF via CAREER award AST-2045928 and grant AST-2107772; NASA ATP grant 80NSSC20K0513; HST grants AR-15809, GO-15902, GO-16273 from STScI.
\bibliography{refs}
\bibliographystyle{aasjournal}
\label{lastpage}

\end{document}